\documentclass{IEEEtran}
\usepackage{amsmath}                 
\usepackage{bm}                      
\usepackage{booktabs}                
\usepackage{commath}                 
\usepackage[inline]{enumitem}        
\usepackage{graphicx}                
\usepackage[pdftex]{hyperref}        
\usepackage{multirow}                
\usepackage{stfloats}                
\usepackage{subcaption}              
\usepackage{tikz}                    
\usepackage{xcolor}                  
\usepackage{xspace}                  

\newcommand{\eg}{\textit{e.g.}\xspace}
\newcommand{\ie}{\textit{i.e.}\xspace}
\newcommand{\Fig}{Fig.\xspace}
\newcommand{\Tab}{Table\xspace}
\newcommand{\gmsh}{\textit{Gmsh}\xspace}
\newcommand{\deriv}[2]{\frac{\partial{}#1}{\partial{}#2}}
\newcommand{\damping}{d}
\newcommand{\M}{M}
\renewcommand{\H}{\vec{H}}
\newcommand{\B}{\vec{B}}
\newcommand{\U}{\vec{u}}
\newcommand{\A}{\vec{A}}
\newcommand{\sP}{\phi}
\newcommand{\dd}{\mathrm{d}}
\newcommand{\Flx}{\mathit{\Phi}}
\newcommand{\angP}{\varphi}
\renewcommand{\S}{\mathcal{S}_\Flx}
\newcommand{\G}{\mathit{\Gamma}_\Flx}
\newcommand{\D}{\mathit{\Omega}}
\newcommand{\Dccc}{\mathit{\Omega}_{\text{ccc}}}
\newcommand{\Dair}{\mathit{\Omega}_{\text{air}}}
\newcommand{\Di}{\mathit{\Gamma}_I}
\DeclareMathOperator{\Curl}{curl}
\DeclareMathOperator{\Grad}{grad}

\newlist{myenum*}{enumerate*}{1}
\setlist[myenum*]{label=\itshape\roman*\upshape)}

\newcommand\copyrighttext{
  \footnotesize
  \textcopyright 2022 IEEE. Personal use of this material is permitted.
  Permission from IEEE must be obtained for all other uses,
  in any current or future media,
  including reprinting/republishing this material for advertising
  or promotional purposes, creating new collective works,
  for resale or redistribution to servers or lists,
  or reuse of any copyrighted component of this work in other works.
}
\newcommand\copyrightnotice{
  \begin{tikzpicture}[remember picture, overlay]
    \node[anchor=south] at (current page.south)
      {\fbox{\parbox{\dimexpr\textwidth-\fboxsep-\fboxrule\relax}
                    {\copyrighttext}}};
  \end{tikzpicture}
}

\newcommand\referencetext{
  \footnotesize
  This article has been accepted for publication
  in IEEE Transactions on Applied Superconductivity.
  This is the author's version which has not been fully edited and
  content may change prior to final publication.
  Citation information: DOI \href{https://doi.org/10.1109/TASC.2022.3206607}
                                 {10.1109/TASC.2022.3206607}
}
\newcommand\referencenotice{
  \begin{tikzpicture}[remember picture, overlay]
    \node[anchor=north] at (current page.north)
      {\parbox{\dimexpr\textwidth-\fboxsep-\fboxrule\relax}
              {\referencetext}};
  \end{tikzpicture}
}

\begin{document}
\title{Influence of mechanical deformations
  on the performance of a coaxial shield for a cryogenic current comparator}

\author{Nicolas~Marsic, Wolfgang~F.~O.~M\"uller,
  Volker~Tympel, Thomas~St\"ohlker, Max~Stapelfeld, Frank~Schmidl,
  Matthias~Schmelz, Vyacheslav~Zakosarenko, Ronny~Stolz,
  David~Haider, Thomas~Sieber, Marcus~Schwickert,
  and Herbert~De~Gersem
  \thanks{Manuscript received xxxxxxxxxxxxxxxxxx; accepted xxxxxxxxxxxxxxxxx.
    Date of publication xxxxxxxxxxxxxx; date of current version xxxxxxxxxxxxxx.
    This project is supported
    by the German Bundesministerium f\"ur Bildung und Forschung
    as the project BMBF-05P18RDRB1
    and the work of Nicolas Marsic is also supported by the Graduate School CE
    within the Centre for Computational Engineering
    at Technische Universit\"at Darmstadt.
    (Corresponding author: Nicolas Marsic.)
  }
  \thanks{N.~Marsic, W.~F.~O.~M\"uller and H.~De~Gersem
    are with the Institute for Accelerator Science and Electromagnetic Fields,
    Technische Universit\"at Darmstadt, 64289 Darmstadt, Germany
    (e-mail:~marsic@temf.tu-darmstadt.de);
    V.~Tympel and T.~St\"ohlker are with the Helmholtz Institute Jena,
    07743 Jena, Germany;
    M.~Stapelfeld and F.~Schmidl are with the Institute for Solid State Physics,
    Friedrich-Schiller-University Jena, 07745 Jena, Germany;
    M.~Schmelz, V.~Zakosarenko and R.~Stolz are with the
    Leibniz Institute of Photonic Technology, 07745 Jena, Germany;
    D.~Haider, T.~Sieber, M.~Schwickert and T.~St\"ohlker are with the
    GSI Helmholtz Centre for Heavy Ion Research, 64291 Darmstadt, Germany;
    V.~Zakosarenko is also with Supracon AG, 07751 Jena, Germany and
    T.~St\"ohlker is also with the
    Institute for Optics and Quantum Electronics,
    Friedrich-Schiller-University Jena, 07745 Jena, Germany.
  }
}
\maketitle
\referencenotice
\copyrightnotice

\begin{abstract}
  This paper studies the impact of mechanical deformations on the performance of
  a coaxial-type cryogenic current comparator (CCC).
  Such deformations may become a concern as the size of the CCC increases
  (\eg when used as a diagnostic device
  in a particle accelerator facility involving beamlines with a large diameter).
  In addition to static deformations,
  this paper also discusses the effect of mechanical vibrations
  on the CCC performance.
\end{abstract}
\begin{IEEEkeywords}
  Cryogenic current comparator,
  current measurement,
  finite element analysis,
  magnetic shielding,
  particle beam measurements.
\end{IEEEkeywords}

\section{Introduction}
\begin{figure}[ht]
  \centering
  \includegraphics{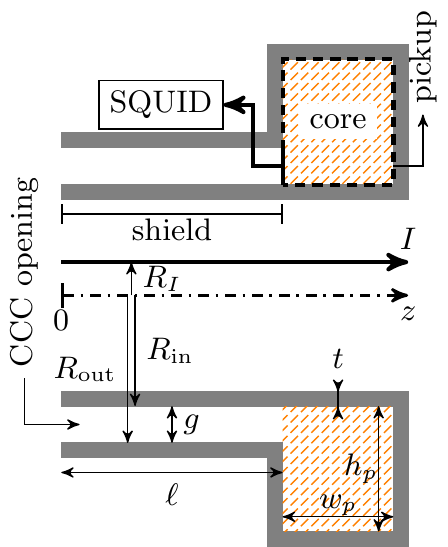}
  \caption{Coaxial-type CCC
    (two-dimensional meridian schematics with symmetry axis);
    in this variant, the CCC itself is part of the pickup loop
    (represented by the thick black dashed lines)
    and contains optionally a highly permeable magnetic core.}
  \label{fig:ccc}
\end{figure}

\IEEEPARstart{T}{he} cryogenic current comparator (CCC) is nowadays
among the most sensitive instruments
to measure very low-amplitude electric currents with high accuracy.
This often cylindrically shaped apparatus contains a superconducting shield
separating all magnetic induction field components
from a superconducting quantum interference
device (SQUID)~\cite{Harvey1972, Williams2011, Clarke2004},
except the azimuthal field component attributed to the current through its bore.
This component is magnetically coupled with a pickup loop
and (optionally) a highly permeable core.
Depending on the geometry of the shield and on the location of the pickup coil,
different variants can be built,
such as the (folded-)coaxial~\cite{Grohmann1976a, Zakosarenko2018, Marsic2019},
the ring~\cite{Grohmann1976b}
or the overlapped tube~\cite{Sullivan1974} CCC shields just to name a few.
Among the applications of those CCCs,
let us mention that overlapped tube variants have been successfully used
in metrological applications,
such as the accurate measurement of resistances for instance
(see \eg \cite{Williams2010a} and references therein).
On the other hand, ring and folded-coaxial CCCs can be used, for example,
in the non-perturbative measurement of low-intensity charged particle beams,
as encountered in particle accelerator facilities~\cite{Zakosarenko2018,
  Fernandes2017}.
In this paper, we will focus on the simplest configuration,
namely the coaxial one, as sketched in \Fig~\ref{fig:ccc}.

The keystone of the CCC's exceptional precision is its
\emph{superconducting} shield
which reduces the amplitude of the magnetic induction field (written as $\B$)~\cite{Grohmann1976a},
with the exception of its azimuthal component that is left intact.
More precisely, by assuming a cylindrical coordinate system $(r, z, \angP)$
and by expanding $\B$ in a Fourier series
along the azimuthal direction,
all Fourier modes but the zeroth one will be attenuated by the shield
(when comparing the amplitude of $\B$
at the CCC opening and at the pickup loop).
This property allows a measurement which
is \emph{quasi independent from the position of the current carrying path through the CCC}.
Formally, let us assume a current $I$ flowing through
a one-dimensional wire located at a radial coordinate $R_I$,
as shown in \Fig~\ref{fig:ccc}.
Then, the \emph{mutual inductance}
\begin{equation}
  \label{eq:M}
  \M(R_I) = \frac{\Flx(R_I)}{I},
\end{equation}
where $\Flx(R_I)$ is the magnetic flux passing through the pickup loop
for a given $R_I$,
\emph{is quasi constant with respect to $R_I$}.
Evidently, $\Flx$ does not depend solely upon $R_I$,
but also on the relative position of the pickup loop with respect to the wire.
This aspect will be further discussed in section~\ref{sec:methodology:pickup}.

The damping of the CCC shield can be mathematically proven
and quantified when considering a geometry with an \emph{axial symmetry}
(\eg see~\cite{Grohmann1976a} or~\cite{Marsic2019}).
However, in practice, mechanical deformations break this symmetry.
Their impact may become non-negligible when considering devices with a large radius and weight.
Moreover, when considering such devices,
low-frequency mechanical vibrations
are triggered by the acoustic environment,
increasing thus the background noise sensed by the whole CCC~\cite{Seidel2018,Tympel2018}.
Therefore, in order to tackle the aforementioned problems, this paper aims at
\begin{myenum*}
\item presenting a numerical framework for quantifying the impact
  of \emph{static} mechanical deformations on the coaxial CCC shield performance;
\item determining whether \emph{static} mechanical deformations impact
  significantly the coaxial CCC shield noise performance and
\item motivating a \emph{quasi-static} approach for studying
  to the impact of mechanical \emph{vibrations}.
\end{myenum*}

This paper is organized as follows.
Section~\ref{sec:num} presents the numerical toolchain
and the mathematical framework for simulating deformed CCCs.
Afterwards, the quantity of interest used for assessing
the performance of the CCC shield is discussed in section~\ref{sec:qoi}.
Subsequently, the numerical methodology used to compute this quantity is exposed
in section~\ref{sec:methodology}
and the simulation results are further presented and discussed
in section~\ref{sec:results}.
Finally, the case of dynamic problems is treated in section~\ref{sec:dyn}
and conclusions are drawn in section~\ref{sec:conclusions}.

\section{Numerical Toolchain for Static Deformations}
\label{sec:num}
In order to assess the CCC performance when undergoing static deformations,
a stepwise approach is employed:
\begin{myenum*}
\item determine a deformed configuration
  starting from the undeformed coaxial-type CCC shown in \Fig~\ref{fig:ccc}
  (see section~\ref{sec:num:mec}) and
\item compute $\M(R_I)$ for this deformed geometry
  (see sections~\ref{sec:num:mag} and~\ref{sec:num:mesh}).
\end{myenum*}
Let us note that a \emph{coreless} variant will be considered
throughout this work, with the exception of section~\ref{sec:results:other}
where the case of CCCs with a magnetic core is briefly discussed.
Indeed, within a coreless framework, the numerical setting can be simplified
by integrating the CCC itself in the pickup loop
(see section~\ref{sec:num:mag} for more details),
as shown in Fig~\ref{fig:ccc}.
This variant was selected
in the folded-coaxial CCC shield presented in~\cite{Zakosarenko2018}.

\begin{figure*}[t]
  \centering
  \subcaptionbox{Mode pattern~I\\(bending).\label{fig:mode:1}}{
    \includegraphics[width=3.25cm]{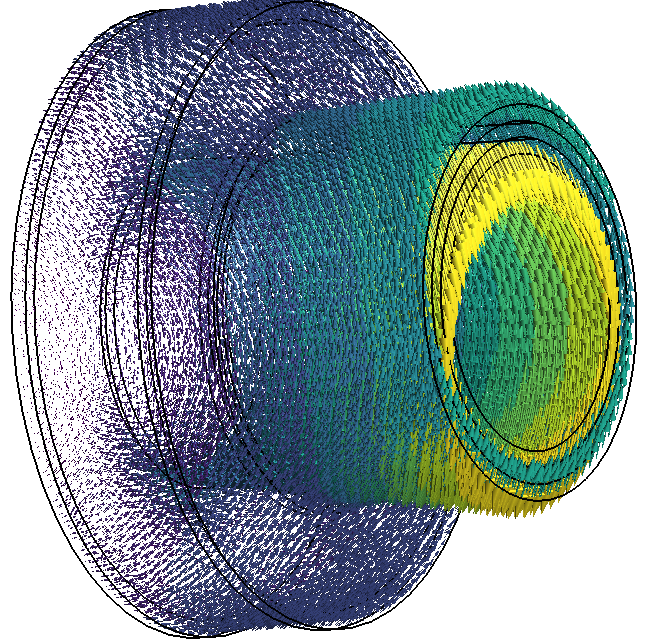}}
  \subcaptionbox{Mode pattern~II\\(axial traction).\label{fig:mode:2}}{
    \includegraphics[width=3.25cm]{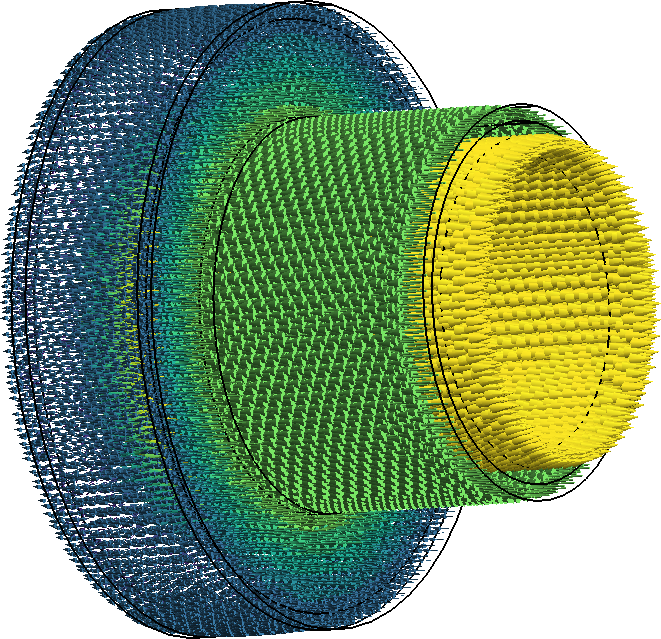}}
  \subcaptionbox{Mode pattern~III\\(radial compression /\\ inner tube).\label{fig:mode:3}}{
    \includegraphics[width=3.25cm]{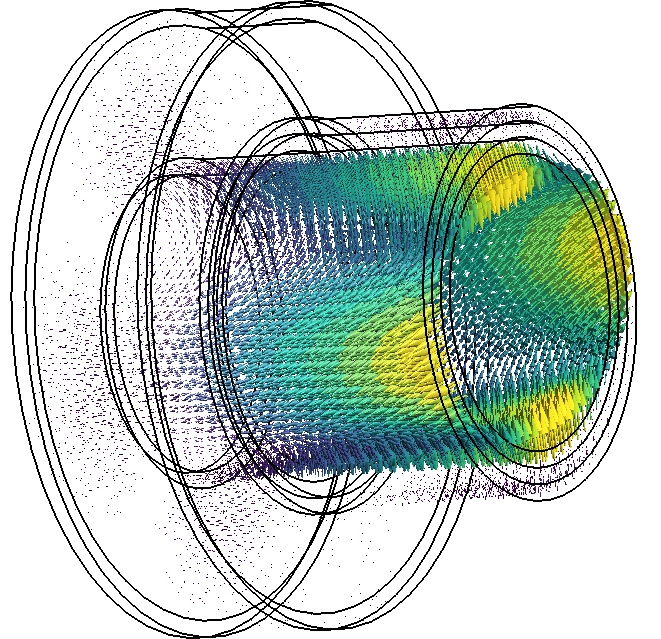}}
  \subcaptionbox{Mode pattern~IV\\(radial compression /\\ outer tube).\label{fig:mode:4}}{
    \includegraphics[width=3.25cm]{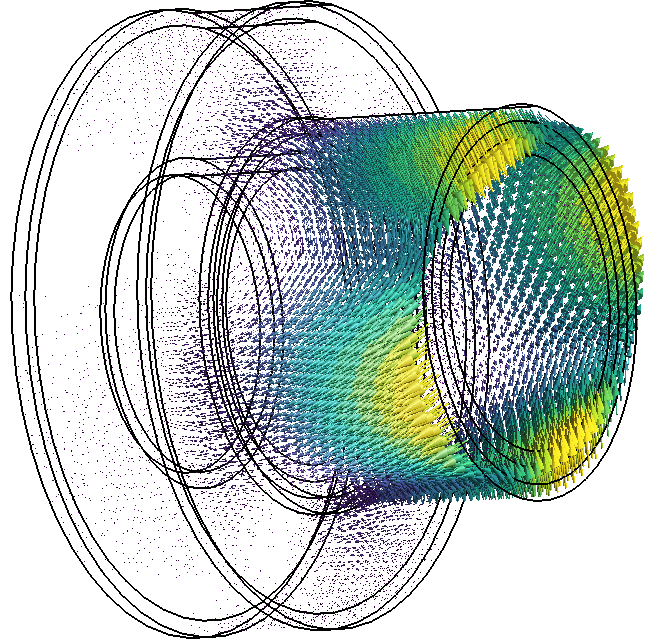}}
  \subcaptionbox{Mode pattern~V\\(torsion).\label{fig:mode:5}}{
    \includegraphics[width=3.25cm]{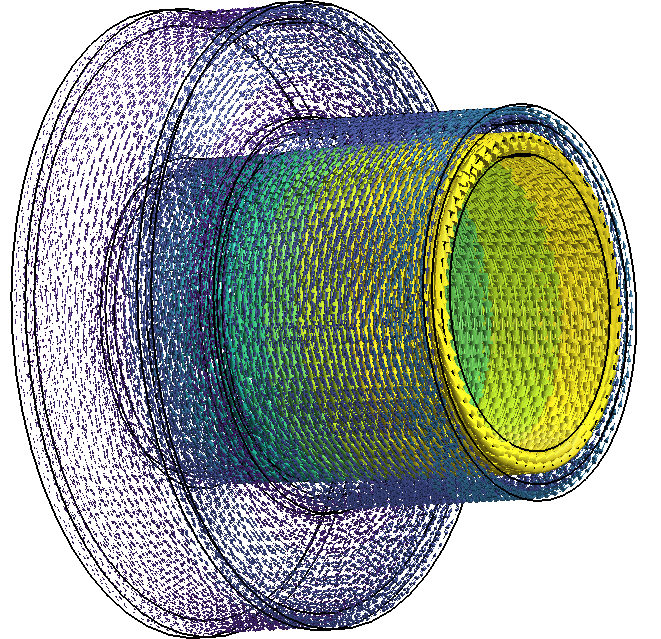}}
  \caption{Normalized mode patterns:
    brightest (resp. darkest) colour for $\norm{\U}=1$
    (resp. $\norm{\U}=0$).}
  \label{fig:mode}
\end{figure*}

\subsection{Mechanical Problem and Mesh Deformation}
\label{sec:num:mec}
Let us start with the mechanical problem:
in order to determine a deformed geometry,
we propose in this work to exploit
the \emph{normal mechanical modes} of the CCC.
More precisely, we consider
the modes of an unclamped CCC (\ie with all six rigid body motions allowed)
free of pressure/traction at its boundary.
This approach is obviously computationally expensive,
but
\begin{myenum*}
\item provides meaningful deformations patterns,
  while considering simple boundary conditions and
\item brings a helpful insight
  into the \emph{dynamical} performance of the CCC,
  as discussed further in section~\ref{sec:dyn}.
\end{myenum*}

The mechanical modes are obtained by discretizing
the time-harmonic elastodynamic equation,
in terms of the displacement field $\U$,
with a second-order finite element (FE) method~\cite{Geradin2015}.
In order to accurately discretize the axisymmetric CCC geometry,
while keeping the number mesh elements as low as possible,
a second-order \emph{curved} mesh is used as well.

From the mode patterns determined by the mechanical solver,
the most important low frequency modes are selected as shown in \Fig~\ref{fig:mode}.
Afterwards, a new set of meshes is generated by deforming the original
axisymmetric CCC mesh according to each of the mode patterns.
Regarding the amplitude of the deformation,
each mode is rescaled such that $\max\norm{\U} = g/s$,
where $g$ is the CCC's air gap (see \Fig~\ref{fig:ccc})
and $s$ a real-valued scaling factor
(see section~\ref{sec:methodology:mec} for more details).
Ultimately, this rescaled mode pattern is employed for deforming the mesh.

For reasons of completeness, let us mention that,
despite the aforementioned advantages,
the computation of the normal mechanical modes is inconvenient in one aspect:
it cannot determine the \emph{amplitude} of a deformation,
as an eigenmode is defined up to a scaling factor.
Such computations can become relevant when assessing that a deformed shield
does not exhibit local short circuits
(\ie contact(s) between different parts of the shield that were not present
in the undeformed configuration).
In this regard, driven simulations (\ie with an actual excitation field)
can be carried out instead (see \eg~\cite{Geradin2015}).
The impact of such short circuits is out of the scope of this work
and is not further considered.

\begin{figure*}[t]
  \centering
  \subcaptionbox{Undeformed CCC\\(mechanical problem).\label{fig:mesh:mec:udf}}{
    \includegraphics[width=5.5cm]{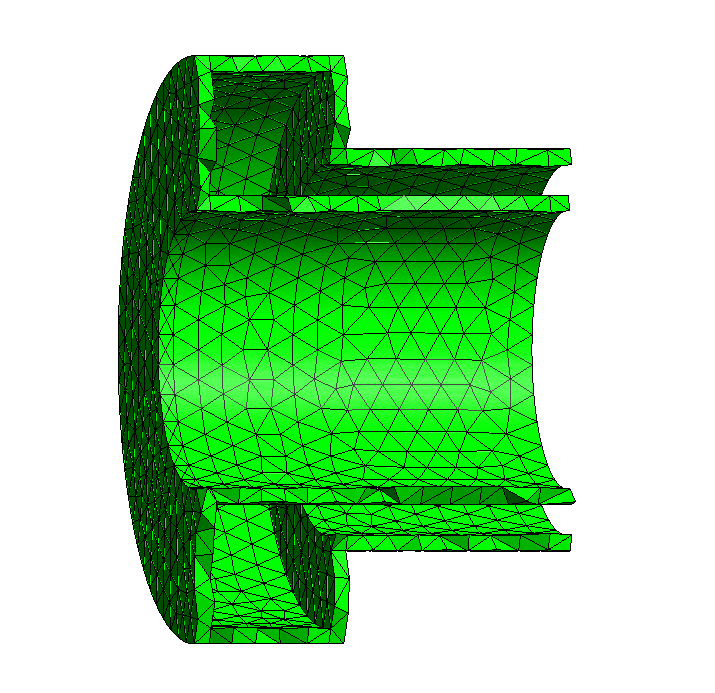}}
  \subcaptionbox{Deformed CCC $\Dccc$\\(mechanical problem).\label{fig:mesh:mec:def}}{
    \includegraphics[width=5.5cm]{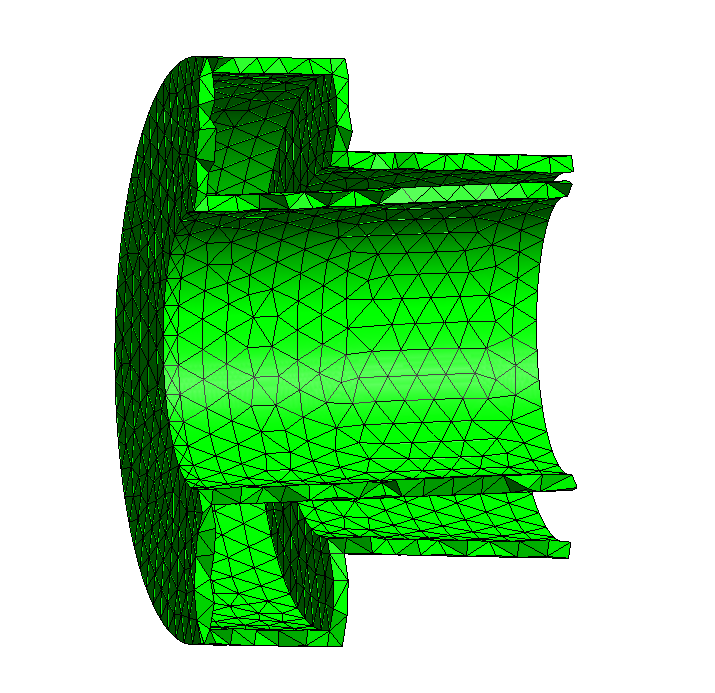}}
  \subcaptionbox{Air domain $\Dair$\\(magnetostatic problem).\label{fig:mesh:mag}}{
    \includegraphics[width=5.5cm]{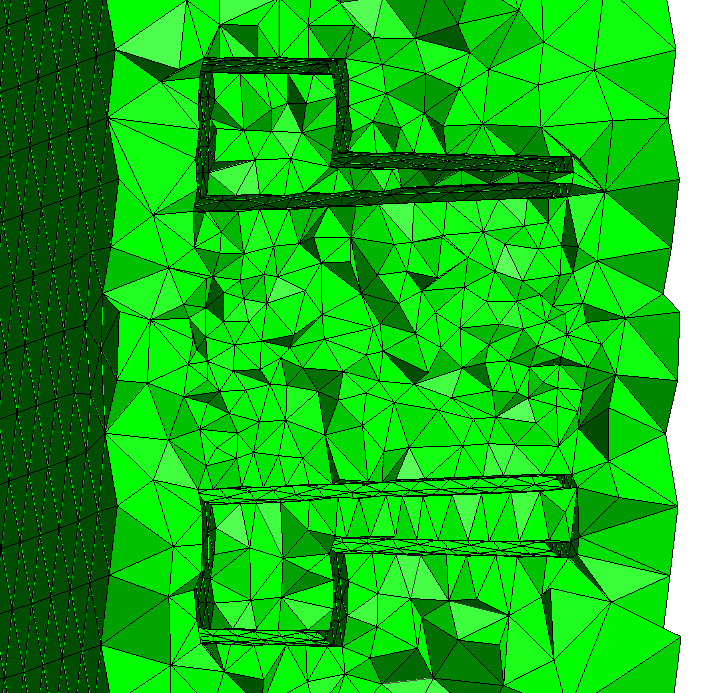}}
  \caption{Finite element meshes
    (exaggerated air gap, exaggerated deformation, meridian view).}
  \label{fig:mesh}
\end{figure*}

\subsection{Magnetostatic Problem}
\label{sec:num:mag}
The magnetic computations can now be carried out on the deformed CCC geometries.
Since the problem considered here is static,
the \emph{magnetostatic} approximation is clearly the appropriate choice.
However, one question remains:
which of the vector-potential FE formulation\footnote{Which is stated
  in terms of $\A$ defined as $\B = \Curl{\A}$.}
or scalar-potential FE formulation\footnote{Which is stated
  in terms of $\sP$ defined as $\H = -\Grad{\sP}$,
  where $\H$ is the magnetic field.}
is the most appropriate~\cite{Bossavit1998a}?
In order to answer this question,
the flux coupling with the SQUID must be first discussed.

When assuming a perfect flux transformer,
the SQUID will measure the magnetic induction flux
passing through the pickup area.
Evidently, if the CCC undergoes a deformation, this pickup area is modified.
However, determining the new pickup surface $\S$ is a difficult task,
\emph{since the deformed geometry is known only
  via a discrete mesh representation}
and not via a computer-aided design (CAD) model.
Nonetheless, when using the \emph{vector}-potential formulation,
this surface must not be known since
\begin{equation}
  \label{eq:phi}
  \Flx = \int_{\S} \B\cdot\dd\vec{n}
       = \int_{\partial\S} \A\cdot\dd\vec{t},
\end{equation}
where $\A$ is the vector-potential (\ie such that $\B = \Curl{\A}$),
$\vec{n}$ is the unit vector normal to $\S$
and $\vec{t}$ is the unit vector tangent to $\partial\S$
(the boundary of $\S$).
In other words, only the \emph{boundary} of the pickup surface needs to be known
when using the vector-potential.
Additionally,
\emph{by assuming that the CCC itself is part of the pickup loop,
  and since the CCC is made of an ideal superconducting material,
  the only portion of $\partial\S$ where $\A\neq{}0$
  is in the air gap}
(\ie the plain line part of the pickup loop in \Fig~\ref{fig:ccc}).
Therefore, by selecting a vector-potential formulation,
the problem of determining $\S$ reduces to the easier task
of finding an appropriate curve $\G$ in the air gap
defining \emph{implicitly}~$\S$.
A simple approach for determining this curve is presented in the next paragraph.

Technically, $\G$ is the path followed
by the wires of the flux transformer between the pickup loop and the SQUID.
However, to avoid the modelling of such a complex path,
and since this study focuses on the CCC \emph{shield},
we propose to define $\G$ as follows: \emph{a straight line segment}
defined by
\begin{myenum*}
\item the plane of constant axial coordinate $z=\ell$
  (where $\ell$ is the shield length);
\item a plane of constant azimuth~$\angP$ and
\item the endpoints located at the radial positions
  $r=R_\text{in}$ (CCC inner radius) and $r=R_\text{out}$ (CCC outer radius),
\end{myenum*}
as shown in \Fig~\ref{fig:ccc} (plain black line segment).
Let us mention that this definition holds for the \emph{undeformed} CCC.
When considering a deformed configuration,
we simply assume that $\G$ remains a straight line segment
anchored at the same endpoints
(whose coordinates have been modified by the deformation).

Now that the use of a vector-potential approach has been motivated,
the magnetostatic problem can be formulated.
By defining the computational domain as
$\D = \Dair\cup\Dccc$, where $\Dair$ (resp. $\Dccc$) refers to
the air surrounding the CCC (resp. the CCC itself),
and by designating the current carrying wire as $\Di$,
the magnetostatic problem reads (in a weak sense):
\begin{align}
  & \text{for all}~\A^\prime\in{}H_0(\Curl, \Dair),~%
    \text{find}~\A\in{}H_0(\Curl, \Dair)~\text{s.t.}\nonumber\\
  & \int_{\Dair}\nu_0\Curl{\A}\cdot\Curl{\A^\prime}\dd\Dair
    =
    \int_{\Di}I\vec{t}\cdot\A^\prime\dd\Di, \label{eq:ms}
\end{align}
since $\A = 0$ in $\Dccc$
(as we assumed the CCC to be made of an \emph{ideal} superconducting material)
and where
\begin{myenum*}
\item $\nu_0$ is the magnetic reluctivity of vacuum,
\item $\vec{t}$ is the unit vector tangent to $\Di$, and
\item $H_0(\Curl,\Dair)$
  is the set of curl-conforming functions defined over $\Dair$
  exhibiting a vanishing tangential component
  on $\partial\Dair$~\cite{Bossavit1998a}.
\end{myenum*}
This variational formulation is discretized
with a second-order FE method and
the \emph{unbounded} domain $\Dair$ is \emph{truncated}
with a shell transformation~\cite{Henrotte1999}.
Additionally, let us mention that the discretized system
is solved with a direct method, requiring thus a gauge condition:
in this work, a spanning tree gauge is exploited~\cite{Dular1995}
to minimise the size of the discrete system.

\subsection{Mesh for the Magnetostatic Problem}
\label{sec:num:mesh}
As already explained in section~\ref{sec:num:mec},
a description of $\Dccc$ (\ie the deformed CCC geometry) is available
in the form of a discrete mesh.
Therefore, a \emph{hybrid} description of $\Dair$ can be obtained, where
\begin{myenum*}
\item its inner boundary is \emph{discretely} defined
  by the facet elements of $\partial\Dccc$ and
\item its volume and outer boundary are defined by a classical CAD model.
\end{myenum*}
In practice, $\Dair$ is a cylindrical domain truncated with a cylindrical shell.

Such hybrid descriptions are possible in the software framework
offered by \gmsh~\cite{Geuzaine2009}.
However, since \emph{curved} meshes are considered in this work,
no guarantee exists (at the time of writing) that the final mesh
obtained from such an hybrid description
will be valid (\ie with an invertible Jacobian matrix).
For this reason, the mesh validity must be checked
\emph{a posteriori}~\cite{Johnen2013}.
In practice, no invalid elements where generated
when using the ``Frontal-Delaunay''~\cite{Rebay1993} (2D mesh)
and the ``HXT''~\cite{Marot2019} (3D mesh) algorithms of \gmsh.
To conclude this section, and for illustration purposes,
both meshes (\ie for $\Dccc$ and $\Dair$) are depicted in \Fig~\ref{fig:mesh}.

\subsection{Software Implementation}
Before concluding this section,
let us mention that the above presented toolchain
is implemented using a homemade FE library\footnote{Available at:
  \url{https://gitlab.onelab.info/gmsh/small_fem}}.
Concerning the mesh related aspects, the software \gmsh is used,
as previously mentioned.

\section{Quantity of Interest}
\label{sec:qoi}
In order to assess the CCC shield performance,
the dependency of $\M(R_I)$, as defined in~\eqref{eq:M}, with respect to $R_I$
is determined,
requiring thus the computation of $\deriv{\M}{R_I}$.
However, as different geometrical configurations will be considered,
we will prefer the dimensionless quantity
\begin{equation}
  \label{eq:eta}
  \eta(R_I) = \deriv{\M}{R_I}\frac{R_\text{in}}{\M_0},
\end{equation}
where $\M_0 = \M(0)$.
Evidently, \emph{low} values of $\eta$ are associated
with \emph{good} position independence properties of the shield.

Additionally,
let us recall that when considering \emph{undeformed} coaxial-type CCCs,
the shield performance is classically characterised
by the dimensionless quantity~\cite{Grohmann1976a}:
\begin{equation}
  \label{eq:d}
  \damping = \exp\frac{\ell}{R_\text{in}},
\end{equation}
where $\ell$ is the length of the shield.
It is important to stress that the above expression relies on the assumption
of an \emph{axisymmetric} shield geometry.
Therefore, $\damping$ cannot be \emph{a priori} used to assess the performance
of a \emph{deformed} shield.
Nonetheless,
and by anticipating the discussion of section~\ref{sec:results},
this simple criterion remains sharp, even in a deformed framework.

While this paper focuses on the shield \emph{itself},
let us briefly discuss the dependence between
the change of magnetic flux in the SQUID of the CCC
and the current $I$ to be measured.
This can be achieved by following for instance the magnetic circuit
approach discussed in~\cite{Zakosarenko2018}\footnote{See equation~(2)
  in~\cite{Zakosarenko2018},
  where $M_A$ is the same as $\M$ in the current work.}.
It is however important to stress that this requires not only
the mutual inductance $\M$
but also the \emph{self} inductance $L$ of the pickup coil.
The latter can be determined
\begin{myenum*}
\item by setting a test current $I_t$ (say $1$~A) in the pickup loop,
\item by setting the current to be measured $I$ to zero,
\item by carrying out the magnetostatic computation for this setting and
\item by extracting $L$ from $\A$.
\end{myenum*}
It is important to stress that if more than one pickup loop is considered
(see below), this procedure must be carried out for \emph{each} loop
(\ie the test current is applied to only one loop at a time).

\section{Numerical Methodology}
\label{sec:methodology}
By exploiting the numerical toolchain presented in section~\ref{sec:num},
the quantity $\M(R_I)$ can be computed
on both geometries: the undeformed and deformed ones.
Afterwards, by evaluating $\M$ for different values of $R_I$,
the performance criterion $\eta$ can be approximated
with a \emph{finite difference} scheme.
Nonetheless, before starting the simulations,
a few practical questions remain to be answered:
\begin{myenum*}
\item the angular position of the pickup surface $\S$
  with respect to the current carrying wire;
\item the geometric parameters of the undeformed CCC; and
\item the mode pattern used for the deformation and its amplitude.
\end{myenum*}
In the subsequent subsections, those three topics are addressed.

\subsection{Pickup Surfaces and Current Carrying Wire}
\label{sec:methodology:pickup}
Obviously, the quantity $\M(R_I)$ is primarily impacted by the
relative position of the current carrying wire
with respect to the pickup surface $\S$.
Therefore,
and in order to draw the most general picture of the CCC shield performance,
different pickup surfaces $\S$ must be considered.
Let us recall that since $\S$ is defined implicitly via $\G$
(see section~\ref{sec:num:mag}),
we need hence to consider
\emph{multiple paths $\mathit{\G}$ associated with different angular positions $\mathit{\angP}$}.

\begin{figure}[ht]
  \centering
  \includegraphics{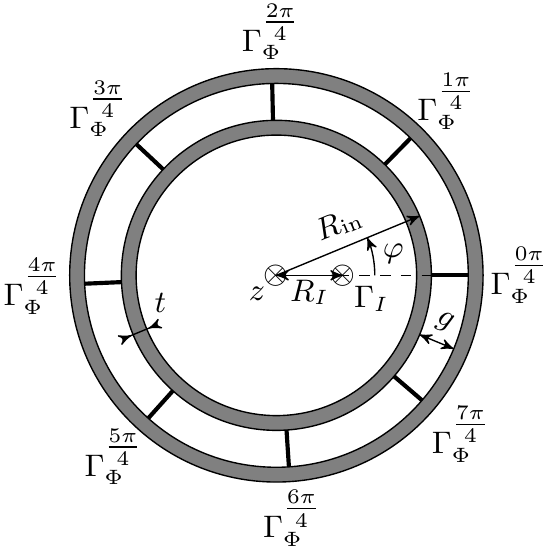}
  \caption{Location of the pickup surfaces $\S^\angP$
    (defined implicitely via $\G^\angP$)
    with respect to the current carrying wire $\Di$
    (two-dimensional axial view, the entire pickup region is not shown).}
  \label{fig:Sphi}
\end{figure}

In particular, we consider in this work eight different angular positions
which are uniformly distributed around the CCC, as illustrated by \Fig~\ref{fig:Sphi}.
Concerning the nomenclature, and in order to distinguish the different surfaces,
the azimuthal position $\angP$ of $\S$ (or $\G$)
will be specified as $\S^\angP$ (or $\G^\angP$).
This typographic convention is furthermore extended to $\M(R_I)$ and $\eta$,
thus $\M^\frac{\pi}{4}(R_I)$ refers to the inductance associated to the
pickup surface located at $\angP=\frac{\pi}{4}$.
Let us note that the first pickup surface is $\S^0$ (or $\G^0$) by convention.

Concerning the current carrying wire $\Di$,
we arbitrarily restrict its position to the meridian plane $\angP=0$,
and assume that a positive value of $R_I$ corresponds to
a decrease of the distance between $\Di$ and $\S^0$,
as suggested in \Fig~\ref{fig:Sphi}.

\subsection{Geometrical Parameters and Undeformed CCCs}
\label{sec:methodology:geo}
In order to determine the dependence of $\eta$
with respect to geometric factors,
different tuples of $R_\text{in}$, $\ell$ and $g$ will be studied.
However, in order to avoid a too high computational cost,
all possible combinations presented will not be considered.
In particular, the simulations will be divided into two groups,
as shown in \Tab~\ref{tab:geo}.
In this table, $h_p$ refers to the height of the pickup surface and
$w_p$ to the width of the pickup surface (see \Fig~\ref{fig:ccc}).
As those two parameters define only the geometry of
the pickup surface and \emph{not of the CCC shield},
they are held constant and chosen as $h_p=w_p=4$ mm in this work.
For all simulations, the thickness of the superconducting material
is selected as $t=0.5$~mm.
\begin{table}[ht]
  \centering
  \caption{Geometrical parameters of the coaxial CCC shield (values in mm).}
  \label{tab:geo}
  \begin{tabular}{rl}
    \toprule
    \multirow[c]{2}{*}{Parameter set 1}
    & $(R_\text{in}, \ell)\in\{2.5, 5, 10\}\times\{8, 16, 32, 64\}$\\
    & with $g=0.5$, $h_p=4$ and $w_p=4$\\
    \midrule
    \multirow[c]{2}{*}{Parameter set 2}
    & $(g, \ell)\in\{0.25, 0.5, 1\}\times\{8, 16, 32, 64\}$\\
    & with $R_\text{in}=5$, $h_p=4$ and $w_p=4$\\
    \bottomrule
  \end{tabular}
\end{table}

\subsection{Mode Patterns and Deformed CCCs}
\label{sec:methodology:mec}
As already mentioned, we will consider in this work the five mode patterns presented
in \Fig~\ref{fig:mode}.
The material properties of niobium at $4.2$~K,
as discussed in~\cite{Lin2003}, are used.
Let us note that while each CCC exhibits those mode patterns,
their relative ordering (with respect to the associated resonance frequency)
might differ between two geometrical configurations.
The numbering given in \Fig~\ref{fig:mode} is associated with a CCC where
$R_\text{in}=5$ mm, $\ell=8$ mm and $g=0.5$ mm.
For illustration purposes,
this configuration exhibits the following resonance frequencies:
$f_\text{Mode I}  \simeq 31$~Hz,
$f_\text{Mode II} \simeq 37$~Hz,
$f_\text{Mode III}\simeq 42$~Hz,
$f_\text{Mode VI} \simeq 43$~Hz and
$f_\text{Mode V}  \simeq 79$~Hz.
In the more realistic case of the FAIR-CCC\footnote{FAIR stands for
  ``Facility for Antiproton and Ion Research'' and is a future
  (at the time of writing) accelerator facility
  based upon an expansion of the
  GSI Helmholtz Centre for Heavy Ion Research,
  Darmstadt, Germany.}~\cite{Seidel2018},
the fundamental mode has been found at approximately $80$~Hz,
by assuming however that the CCC is clamped at its inner radius.

As discussed in section~\ref{sec:num:mec},
each mode pattern is first rescaled such that $\max\norm{\U} = g/s$
before deforming the CCC mesh accordingly.
In this work, $s$ is selected in the set $s\in\{2, 4, 8\}$.
Before concluding this section,
let us indicate that because of the axial symmetry,
the mode patterns are degenerated along the azimuthal direction.
In order to compare similar patterns,
this degeneracy is lifted by rotating the mode pattern
such that $\max\norm{{\U}}$ is aligned with the angular position $\angP=0$.
Let us also note that a sign inversion might be additionally required.

\section{Results and Discussion}
\label{sec:results}
In this section, the results of the simulations described
in the previous section are discussed.
However, before entering into the heart of this work,
let us first discuss the numerical accuracy of our computations.

\subsection{Numerical Accuracy}
\begin{figure}[ht]
  \centering
  \includegraphics{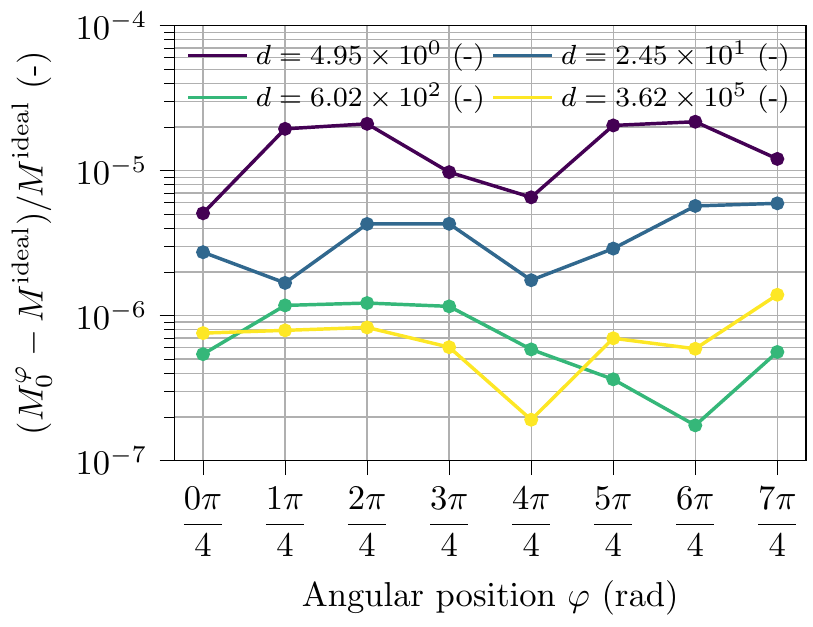}
  \caption{Accuracy of the magnetostatic FE solution
    for the ideal undeformed case
    ($R_{\text{in}}=5$~mm and $g=0.5$~mm).}
  \label{fig:accuracy}
\end{figure}

In order to determine the accuracy of the magnetostatic simulations,
let us remind that in the ideal case,
\ie with the current carrying wire located directly on the symmetry axis and without deformations,
the inductance $\M^{\text{ideal}}$ can be computed with
the formula for toroidal coils
with a rectangular cross section~\cite{Durand1953}:
\begin{equation}
  \label{eq:Ma}
  \M^{\text{ideal}} =
  \frac{I}{\nu_0}\frac{w_p}{2\pi}\ln\frac{R_{\text{in}}+h_p}{R_{\text{in}}},
\end{equation}
since $\B$ is fully azimuthal in this case
and the CCC shield is thus transparent.
While being an idealised setting,
this simpler configuration offers a practical way to estimate the FE error.
In the case of the parameter set
$R_{\text{in}}=5$ mm and $g=0.5$~mm,
the relative error is presented in \Fig~\ref{fig:accuracy}.
Let us note that this figure shows the error for different vales of
$\ell\in\{8, 16, 32, 64\}$~mm.
However, in order to keep this parameter dimensionless,
the coaxial damping $\damping(R_{\text{in}}, \ell)$,
as defined in equation~\eqref{eq:d}, is preferred.
Its systematic use will be motivated in the next subsections.

From the data shown in \Fig~\ref{fig:accuracy},
it is clear that the FE error decreases with the parameter $\damping$
until it reaches a plateau at approximately $10^{-6}$:
it is interesting to notice that the damping properties of the CCC shield are
also expressed at the discretized level.
Let us also indicate that a similar behaviour of the FE error is observed
for each parameter set in \Tab~\ref{tab:geo},
up to a few exceptions associated with a low damping $\damping<50$.
We also observe that the FE error is quasi symmetric with~$\angP$
when $\damping = 4.95$,
but becomes more and more asymmetric as $\damping$ increases.
This behaviour can be explained by the fact that
the FE mesh is not perfectly symmetric.
Therefore, as the numerical error decreases with increasing values of $\damping$,
this asymmetry becomes more visible.
In addition, the accuracy of the linear system solver is not infinite.

Concerning the mechanical simulations,
the validity of the computations is assessed with a mesh refinement analysis
based on the eigenvalues associated with the mode patterns.
However, given the computational cost of such an analysis,
only the smallest geometric configuration is considered
with only three level of refinements.
In this regard,
the maxi\-mum relative change between two levels is smaller than $3~\%$.
Nonetheless, and by anticipating the next subsections,
the impact of the considered mode patterns on $\eta$ is low,
and therefore highly accurate mechanical simulations are not required.

\subsection{Impact of $\damping$ on $\eta^\angP$}
Now that our numerical models are validated,
let us enter the core of this work
and discuss the influence of deformations on $\eta^\angP$.
To begin with, let us investigate the dependency of $\eta^\angP$
with respect to the shield length,
or more precisely the coaxial damping $\damping(R_{\text{in}}, \ell)$.
It is worth stressing that the CCC is \emph{deformed} in this test case,
while the quantity $\damping$ refers to the coaxial damping
in the \emph{undeformed} configuration.

\begin{figure}[ht]
  \centering
  \includegraphics{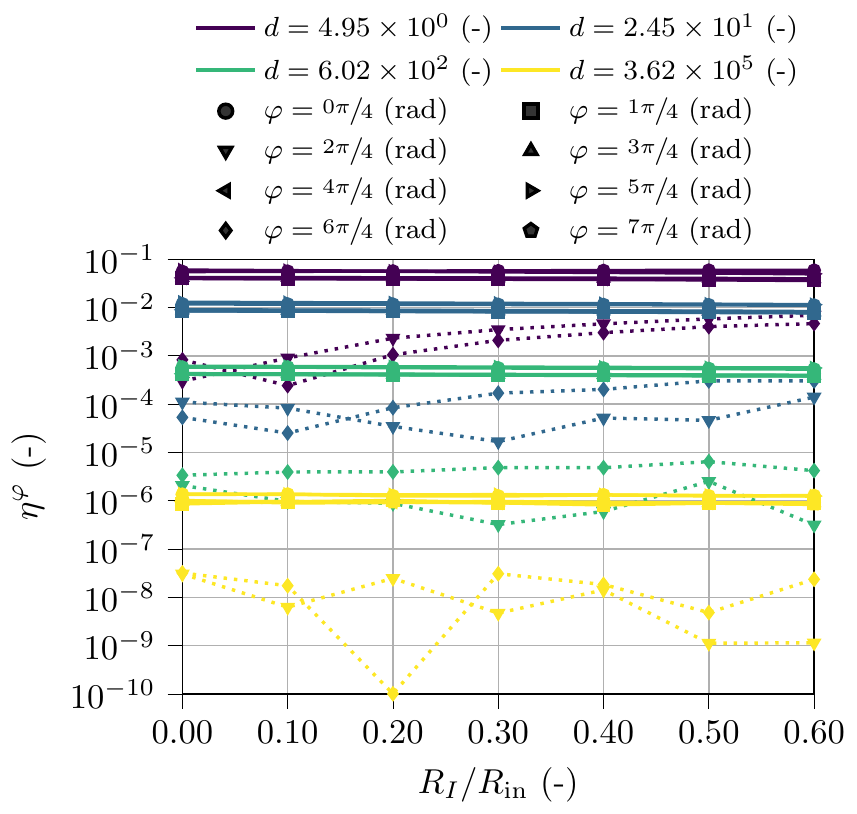}
  \caption{Impact of $\damping$ on $\eta^\angP$
    ($R_{\text{in}}=5$ mm, $g=0.5$ mm,
    mode pattern~I and $s=4$).}
  \label{fig:eta:delta}
\end{figure}

\Fig~\ref{fig:eta:delta} shows the behaviour of $\eta^\angP$
with respect to $R_I/R_{\text{in}}$ for different values of
$\damping$ and $\angP$.
In those data,
\begin{myenum*}
  \item the geometrical parameters are fixed to
    $R_{\text{in}}=5$ mm and $g=0.5$ mm;
  \item the deformation pattern~I is used; and
  \item the deformation scaling factor is fixed to $s=4$.
\end{myenum*}
From those data, it is clear that two angular positions,
namely $\angP=\pi/2$ and $\angP=3\pi/2$,
exhibit a significantly lower value for $\eta^\angP$,
as compared to the other angular positions \emph{for a fixed $\damping$}.
The existence of two particular angular positions can be justified theoretically
in the undeformed case, as demonstrated in appendix~\ref{sec:phieta}.
Furthermore, their precise location at $\angP=\pi/2$ and $\angP=3\pi/2$
is a consequence
of the restriction of the current carrying wire to the plane $\angP=0$,
as also shown in appendix~\ref{sec:phieta}.
In order to improve the readability of the subsequent plots,
those two special angles are omitted from now on.

Additionally, it is obvious from \Fig~\ref{fig:eta:delta} that
the higher the~$\damping$, the lower the~$\eta^\angP$.
In particular, for an increase of $\damping$ by a power of $2$,
$\eta^\angP$ is reduced by a power of $\sim{}2$ as well.
In light of this behaviour,
it is tempting to directly conclude that the \emph{intrinsic} property
of the CCC shield,
\ie the damping of the non-azimuthal components of $\B$
(in the sense of the Fourier decomposition discussed in the introduction),
are kept despite the deformations.
Nonetheless, this hypothesis must be first validated by means of other
numerical experiments, as discussed in the next subsections.

\subsection{Impact of the Geometrical Parameters on $\eta^\angP$}
\begin{figure}[t]
  \centering
  \includegraphics{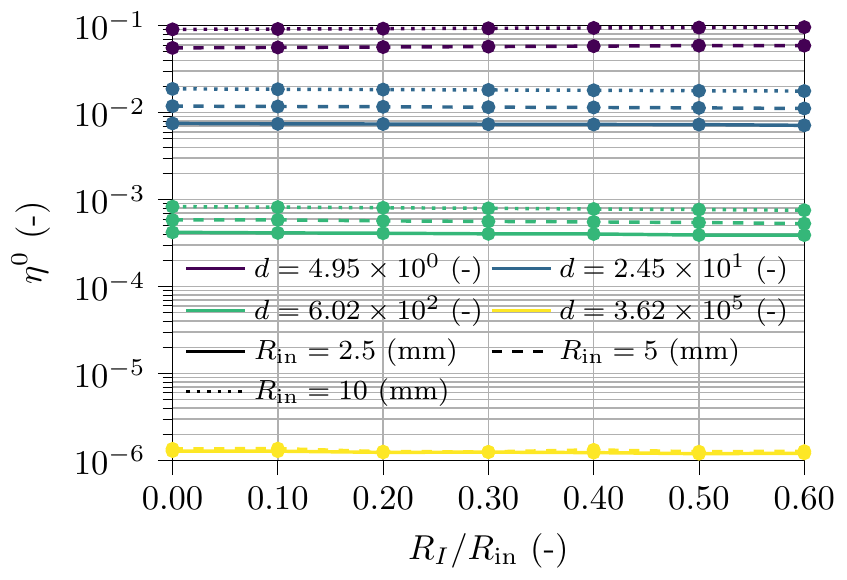}
  \caption{Impact of $R_\text{in}$ on $\eta^0$
    (parameter set 1, mode pattern~I and $s=4$).}
  \label{fig:eta:R}
\end{figure}

In order to validate our previous hypothesis,
let us start with determining the impact of the inner radius
$R_{\text{in}}$ on $\eta^\angP$
for different values of $\damping(R_\text{in}, \ell)$,
which corresponds to the first data set in \Tab~\ref{tab:geo}.
This influence is depicted in \Fig~\ref{fig:eta:R} for the case $\angP=0$.
Let us note that a similar behaviour is observed for the other angles,
but only $\angP=0$ is shown for clarity reasons.
Let us also mention that the combinations
$(R_{\text{in}}, \damping)\in\{(2.5~\text{mm}, 4.95),
(10~\text{mm}, 3.62\times{}10^5)\}$ do not exist.
From the data displayed in \Fig~\ref{fig:eta:R},
it is obvious that,
while $R_{\text{in}}$ affects $\eta^\angP$,
its impact is negligible in regard to $\damping$.

\begin{figure}[t]
  \centering
  \includegraphics{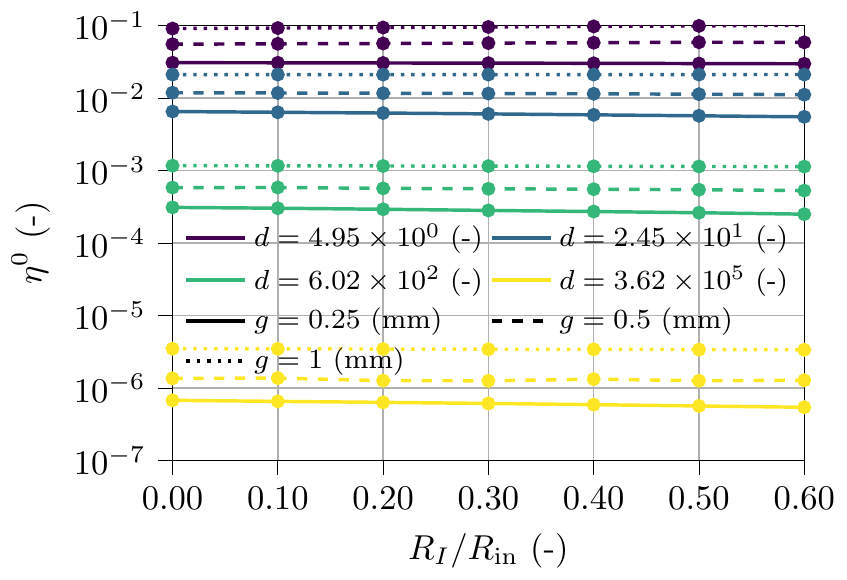}
  \caption{Impact of $g$ on $\eta^0$
    (parameter set 2, mode pattern~I and $s=4$).}
  \label{fig:eta:g}
\end{figure}

In \Fig~\ref{fig:eta:g}, the influence of the air gap size $g$
on $\eta^\angP$ is shown
for different values of $\damping(R_\text{in}=5~\text{mm}, \ell)$.
This case corresponds to the second data set in \Tab~\ref{tab:geo},
and  only the case $\angP=0$ is shown for the same above mentioned reasons.
It is clear from \Fig~\ref{fig:eta:g} that an increase in $g$
leads to an increase in $\eta$.
Let us note that this behaviour is expected,
since larger air gap sizes are known to decrease the coaxial CCC shield performance
even in the ideal undeformed case~\cite{Marsic2019}.

\subsection{Impact of Deformations on  $\eta^\angP$}
\begin{figure}[ht]
  \centering
  \includegraphics{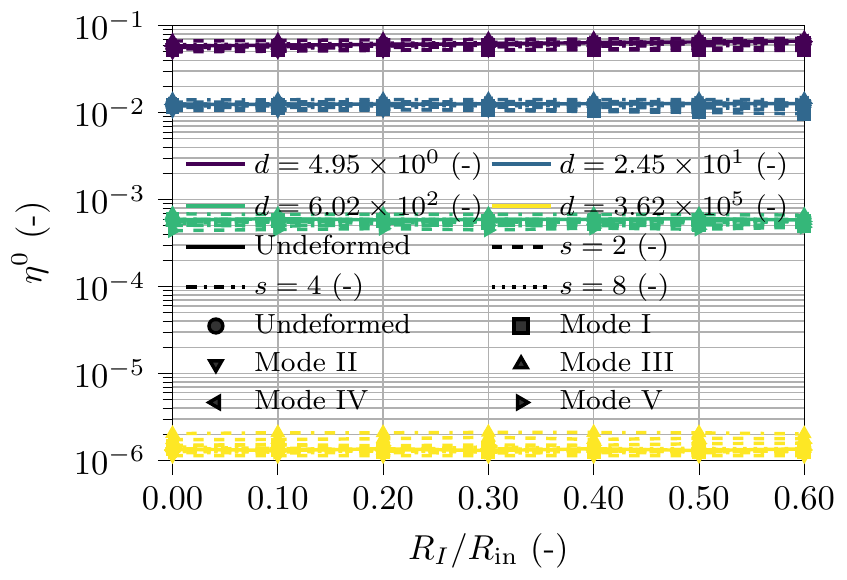}
  \caption{Impact of deformations on $\eta^0$
    (parameter set 1 restricted to $R_{\text{in}}=5$ mm).}
  \label{fig:eta:mode:general}
\end{figure}

Now that the effect of the geometrical parameters are assessed
for a deformed coaxial-type CCC,
let us focus on the impact of the deformations themselves on $\eta^\angP$.
Again, the data presented below are restricted to the case $\angP=0$
for clarity reasons, but the discussion remains valid
for the other angular positions.
As it can be seen in \Fig~\ref{fig:eta:mode:general},
the performance of the coaxial CCC shield
is only mildly affected by the deformations,
when compared to the impact of~$\damping$.
In particular, this figure shows $\eta^0$
for all five deformation patterns
and for all three deformation amplitudes
together with the undeformed configuration.
The geometric parameters are those of the parameters set 1 in \Tab~\ref{tab:geo}
restricted to $R_{\text{in}}=5$~mm.
Nonetheless,
the same conclusion can be drawn for the other values of $R_{\text{in}}$,
which are not shown here for conciseness reasons.

In order to complement the data shown in \Fig~\ref{fig:eta:mode:general},
let us first take the average value of $\eta^0$ over $R_I/R_\text{in}$,
as it is clear from \Fig~\ref{fig:eta:mode:general}
that $\eta^0$ is quasi independent from $R_I/R_\text{in}$,
and denote it by $\overline{\eta^0}$.
We then compute the relative deviation $\varepsilon_{d,m,s}$ for a given value of $\damping$,
a given deformation pattern $m$ and a given value of $s$:
\begin{equation}
  \label{eq:eta:mode:error}
  \varepsilon_{\damping,m,s} = \frac{\abs{\overline{\eta^{0,\text{def}}_{\damping,m,s}}
                                         -\overline{\eta^{0,\text{und}}_{\damping}}}}
                                    {\overline{\eta^{0,\text{und}}_{\damping}}},
\end{equation}
where $\overline{\eta^{0,\text{def}}_{\damping,m,s}}$ is associated
with the deformed geometry
and $\overline{\eta^{0,\text{und}}_{\damping}}$
with the undeformed one\footnote{The undeformed geometry obviously does not
  depend on a deformation pattern $m$ or on the amplitude $s$.}.
Afterwards, we collect in Table~\ref{tab:eta:mode:general}
the maximum value of $\varepsilon_{\damping,m,s}$ for each $\damping$, that is formally
\begin{equation}
  \label{eq:eta:mode:error:max}
  \varepsilon_{\damping}^{\max} = \max_{(m,s)}\varepsilon_{\damping,m,s}.
\end{equation}
From these data,
the relative variations $\varepsilon_{\damping}^{\max}$ appear to be large,
but it must be noted that they are small
compared with the variations induced by $\damping$ itself.
Let us also note that the increase in $\varepsilon_{\damping}^{\max}$ with $\damping$
must be considered with care,
as it might be simply due to the limit of the numerical model
being reached\footnote{This limit can be improved
  by using finer FE meshes and iterative refinement~\cite{Arioli1989}.}.
\begin{table*}[t]
  \centering
  \begin{tabular}{ccccc}
    \toprule
    $\damping$
    & $\varepsilon_{\damping}^{\max}$
    & Associated pair $(m, s)$
    & Associated $\overline{\eta^{0,\text{und}}_{\damping}}$
    & Associated $\overline{\eta^{0,\text{def}}_{\damping,m,s}}$\\
    \midrule
    $4.95\times{}10^0$ & $16$~\% & $(I,   2)$ & $6.22\times{}10^{-2}$ & $5.24\times{}10^{-2}$\\
    $2.45\times{}10^1$ & $16$~\% & $(I,   2)$ & $1.26\times{}10^{-2}$ & $1.06\times{}10^{-2}$\\
    $6.02\times{}10^2$ & $22$~\% & $(IV,  2)$ & $5.91\times{}10^{-4}$ & $4.58\times{}10^{-4}$\\
    $3.62\times{}10^5$ & $33$~\% & $(III, 2)$ & $1.34\times{}10^{-6}$ & $1.78\times{}10^{-6}$\\
    \bottomrule
  \end{tabular}
  \caption{Relative and absolute variations of the shield performance
    (all quantities are dimensionless,
    parameter set 1 restricted to $R_{\text{in}}=5$ mm).}
  \label{tab:eta:mode:general}
\end{table*}

\subsection{Impact of Deformations on  $\M^\angP(0)$}
So far, our discussion was focused solely on $\eta^\angP$,
which reflects how effective the coaxial CCC shield is
in terms of position independence.
While this paper aims primarily at studying this aspect,
the behaviour of $\M$ itself deserves nonetheless a few comments.
As mentioned in the previous subsection,
the value of $\eta$ is only mildly affected by deformations:
it is therefore legitimate to concentrate this subsection
on $\M^\angP(0)=\M^\angP_0$ only.

\begin{figure}[ht]
  \centering
  \includegraphics{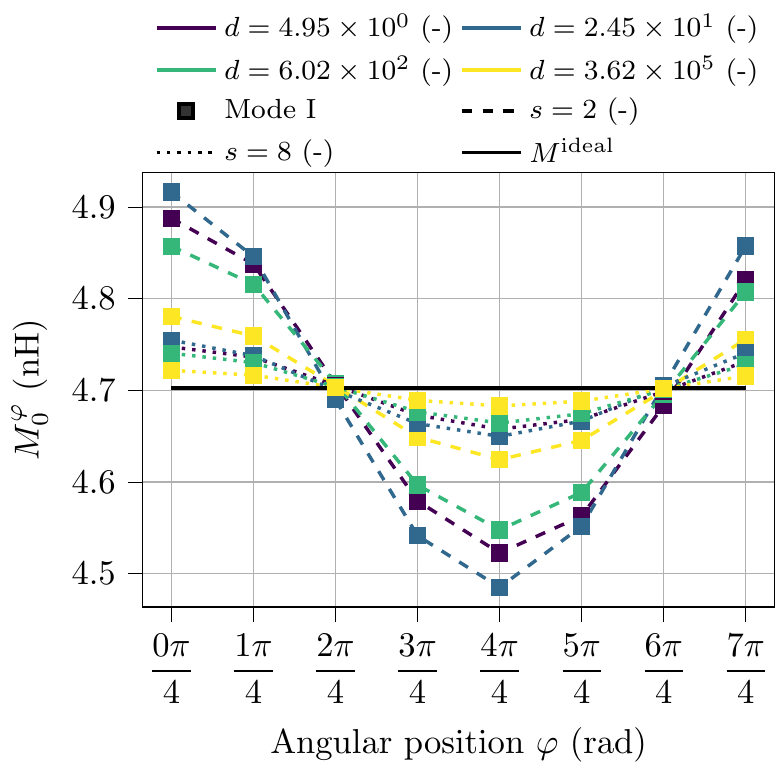}
  \caption{Impact of deformations on $\M^\angP(0)=\M^\angP_0$
    (parameter set~1 restricted to $R_{\text{in}}=5$ mm).}
  \label{fig:M}
\end{figure}

As it can be seen in \Fig~\ref{fig:M},
deformations have clearly a higher impact on $\M$ than $\eta$.
This behaviour is easily explained,
since \emph{the value of $\M$ is directly related to the pickup surface},
which can undergo significant deformations.
It is also unequivocal from \Fig~\ref{fig:M} that deformations have less impact
on $\M_0$ for larger values of $\damping$.
\emph{The reason for this is however unrelated to the magnetic properties
  of the coaxial CCC shield}.
Indeed,
a deformation whose maximum amplitude is located at the tip of the CCC opening
(exactly as the mode patterns in \Fig~\ref{fig:mode})
is impacting less the pickup surface when $\ell$ (and thus $\damping$) is large,
since the distance between the maximum displacement
and the pickup surface is increased.

\subsection{Summary and Additional Comment}
\label{sec:results:other}
We recognized in this section that $\eta$ can be significantly lowered
when $\damping$ is large,
even if some geometrical parameters may worsen its value mildly.
Moreover, we observed that the amplitude of the deformation or the mode pattern
have also a negligible impact on $\eta$ compared to $\damping$.
For all these reasons, it is legitimate to conclude that
\emph{the coaxial CCC shield keeps its intrinsic damping properties,
  even when considering deformations}.
Nonetheless, as a direct consequence of the deformations of the pickup surface,
deformations have a non-negligible impact on $\M$ itself.

It is also worth mentioning that we considered
a coreless variant of the coaxial-type CCC,
where the CCC itself is part of the pickup loop.
In the case of a CCC with a highly permeable magnetic core,
it is however more usual to separate shield and pickup circuit by winding a superconducting coil around the core.
Nonetheless, as this variation does not modify the \emph{shield itself},
our conclusions about $\eta$ and the coaxial CCC shield damping remain unchanged.
Let us also note that CCCs with highly magnetic permeable materials
suffer from temperature dependent drifts and low-frequency noise
caused by flipping of individual magnetic domains
inside the core material~\cite{Zakosarenko2018}.

To conclude this section,
let us recall that our model cannot allow arbitrary deformations.
Indeed, large deformations with respect to the air gap my lead to
new contacts within the shield (see section~\ref{sec:num:mec}).
Those topological changes may lead to drastic modifications
in the shielding properties, but are out of the scope of this paper.

\section{Dynamic Problems}
\label{sec:dyn}
In the previous section, only \emph{static} deformations were investigated.
However,
the CCC can be submitted to mechanical \emph{vibrations} during operation,
especially in the context of particle beam diagnostics.
For this reason, the treatment of dynamic problems is also highly relevant.
From an electromagnetic point of view,
the position independence property of the CCC shield follows
from the ability of superconducting materials
to expel any interior magnetic induction field~\cite{Grohmann1976a}.
This perfect diamagnetism is attributed to shielding currents,
which are in practice well described by the London theory~\cite{London1935},
as CCCs are usually made of thick superconductors
(\eg $3$ mm in the case of the FAIR-CCC~\cite{Seidel2018}
or $1$ mm for the CCC discussed in~\cite{Zakosarenko2018})
held well below the critical temperature.
In this theoretical framework, and by neglecting the displacement currents%
\footnote{Which is legitimate,
  since one can expect the mechanical resonance spectrum
  to be orders of magnitudes lower that the electromagnetic resonance spectrum,
  allowing us therefore to neglect electromagnetic wave phenomena.},
the London equation do not contain any dynamics.
For this reasons, the static electromagnetic properties of the CCC shield remain valid
even in a dynamic context,
allowing us to conclude that
\begin{myenum*}
\item our static analysis can be applied in a dynamic context
  (\emph{quasi-static} framework) and
\item the \emph{intrinsic} performance of the coaxial CCC shield are kept,
  even when undergoing mechanical vibrations,
\end{myenum*}
at least as long as our working hypothesis are met.
In particular, we assumed implicitly that the stress and strain fields
do not impact the electromagnetic properties of the superconducting material.
Let us note that the same holds for the magnetic core
when this one is present,
this latter hypothesis being however quite restrictive in practice.

\section{Conclusions}
\label{sec:conclusions}
Via numerical computations, we showed
that the \emph{intrinsic} position independence property
of the coaxial CCC shield is negligibly affected by mechanical deformations.
Nonetheless, as those deformations may change the size and orientation of the CCC pickup surface,
the coupling inductance $\M$ may be significantly affected.
We furthermore discussed the case of mechanical vibrations,
and showed that the above mentioned conclusions remain valid,
as long as London theory remains valid,
which we expect to be true in practical situations.
Last but not least, let us also mention that the numerical toolchain developed
in this work is not specific to the analysis of CCCs
and can be used to assess the performance
of any deformed superconducting magnetic shield.
Additionally, this approach can be also used to analyse CCCs
with other types of shielding, such as the ring or the overlapped ones.

\section*{Acknowledgement}
The authors would like to express their gratitude to
Ms.~Heike Koch, Mr. Achim Wagner, Mr. Dragos Munteanu and Mr. Christian Schmitt
for the administrative and technical support.

\appendices
\section{Why is $\eta^\angP$ lower
  for $\angP=\pi/2$ and $\angP=3\pi/2$?}
\label{sec:phieta}
As already mentioned in the introduction,
by expanding $\B$ in a Fourier series along the azimuthal direction,
it can be shown that the coaxial CCC shield damps
all Fourier modes but the zeroth one.
In particular, \emph{the higher the Fourier mode,
the higher the attenuation}~\cite{Grohmann1976a}.
In the case of the numerical setting
presented in section~\ref{sec:methodology:pickup},
the source current density $\vec{j}_s$ due to the current carrying wire
can be formally written in a cylindrical coordinate system
as (see \Fig~\ref{fig:Sphi}):
\begin{equation}
  \label{eq:j:delta}
  \vec{j}_s = \frac{I}{R_I}\delta(\angP)\hat{z},
\end{equation}
where $\delta$ is a $2\pi$-periodic \emph{Dirac delta function}
and $\hat{z}$ is the unit vector oriented along the symmetry axis.
It is worth mentioning that this expression is compatible
with the right-hand side of~\eqref{eq:ms} when restricted to $\Di$.

Let us now expand~\eqref{eq:j:delta} in a Fourier series along $\angP$:
\begin{equation}
  \label{eq:j:fourier}
  \vec{j}_s
  =
  \frac{I}{R_I}\left(\frac{1}{2\pi}+
                     \frac{1}{\pi}\sum_{n=1}^\infty\cos(n\angP)\right)\hat{z},
\end{equation}
and define its modal components as
\begin{equation}
  \label{eq:j:n}
  \vec{j}^0_s = \frac{I}{2\pi R_I}\hat{z}%
  \quad\text{and}\quad%
  \vec{j}^n_s = \frac{I}{\pi R_I}\cos(n\angP)\hat{z}~\forall{}n>0.
\end{equation}
According to the aforementioned theory
of undeformed coaxial CCC shields~\cite{Grohmann1976a},
each modal component of $\vec{j}_s$ will be damped,
with $\vec{j}^1_s$ undergoing the lowest damping
(by excluding of course $\vec{j}^0_s$ which is undamped).
Therefore, when considering the angular positions $\angP=\pi/2$
and $\angP=3\pi/2$,
it is clear that $\vec{j}^1_s = 0$.
Thus, the component undergoing the lowest damping is for those angles unexcited,
reducing thereby $\eta^\angP$ by a large factor when compared to $\angP=0$.

\bibliographystyle{ieeetr}
\bibliography{bib.bib}
\end{document}